\input{aipcheck}

\documentclass[
   final            % use final for the camera ready runs
  ]
  {aipproc}

\layoutstyle{6x9}

\begin{document}

\title{
Nuclear Effects on
Generalized Parton Distributions
of {$^3$He }
}

\author{Sergio Scopetta}{
  address={Dipartimento di Fisica, Universit\`a degli Studi
di Perugia, via A. Pascoli
06100 Perugia, Italy and INFN, sezione di Perugia;
\\
Departament de Fisica Te\`orica,
Universitat de Val\`encia, 46100 Burjassot (Val\`encia), Spain
} }

\begin{abstract}
The relevance of measuring
generalized parton distributions (GPDs)
of nuclei is stressed and the unique possibilities
offered by nuclear few body systems are emphasized. 
A realistic microscopic calculation
of the unpolarized quark 
GPD $H_q^3$ of the $^3$He nucleus is reviewed.
Nuclear effects are found to be 
larger than in inclusive deep inelastic scattering,
flavor dependent,
increasing with the momentum
transfer and the asymmetry of the process.
They also depend
on the realistic nuclear potential
chosen to estimate them.
Besides, it is found that 
nuclear GPDs cannot be factorized into a $\Delta^2$-dependent
and a $\Delta^2$-independent term, as suggested
in prescriptions proposed for finite nuclei.
\end{abstract}

\maketitle

%%%%%%%%%%%%%%%%%%%%%%%%%%%%%%%%%%%%%%%%%%%%
%% MAINMATTER
%%%%%%%%%%%%%%%%%%%%%%%%%%%%%%%%%%%%%%%%%%%%

%\section{<A section>}

Generalized Parton Distributions (GPDs) \cite{first} 
parametrize the non-perturbative hadron structure
in hard exclusive 
processes
(for a recent review, 
see, e.g.,  \cite{dpr}),
entering
the long-distance dominated part of
exclusive lepton Deep Inelastic Scattering
(DIS) off hadrons.
In particular, Deeply Virtual Compton Scattering (DVCS),
i.e. the process
$
e H \longrightarrow e' H' \gamma
$ when
$Q^2 \gg m_H^2$,
is one of the the most promising to access GPDs.
Here and in the following,
$Q^2$ is the momentum transfer between the leptons $e$ and $e'$,
and $\Delta^2=(P'-P)^2$ the one between the hadrons $H$ and $H'$,
which have momenta $P$ and $P'$, respectively.
GPDs depend on $\Delta^2$, on the so called
skewness parameter, given by $\xi = \Delta^+ / (P + P')^+$,
and on the fraction of light cone momentum $x$.
The dependence on the scale $Q^2$ will not be discussed here.

Recently, the issue of measuring GPDs for nuclei
has been addressed. In the first paper on this subject
\cite{cano1}, concerning the deuteron,
it has been observed that
the knowledge of GPDs would permit the investigation
of the short light-like distance structure of nuclei, 
and thus the interplay of nucleon and parton degrees of freedom 
in the nuclear wave function.
In standard DIS off a nucleus
with four-momentum $P_A$ and $A$ nucleons of mass $M$,
this information can be accessed in the 
region where
$A x_{Bj} \simeq {Q^2\over 2 M \nu}>1$,
being $x_{Bj}= Q^2/ ( 2 P_A \cdot q )$ and $\nu$
the energy transfer in the laboratory system.
In this region measurements are very difficult, because of 
vanishing cross-sections. As explained in \cite{cano1}, 
the same physics can be accessed
in DVCS at much lower values of $x_{Bj}$.
The usefulness of nuclear GPDs
has been stressed also for finite nuclei in
Refs. \cite{poly}.

The study of GPDs for $^3$He is interesting
for many aspects. 
In fact, $^3$He is a well known nucleus, for which realistic studies 
are possible, so that conventional nuclear effects
can be safely calculated. Strong deviations from the predicted
behavior could be therefore ascribed to exotic effects, such as
the ones of non-nucleonic degrees of freedom, not included in a
realistic wave function.
Besides, $^3$He is extensively used as an effective neutron target.
In fact, the properties of the free neutron are being investigated
through experiments with nuclei, whose data
are analyzed taking nuclear effects properly into account.
Recently, it has been shown that unpolarized 
DIS off three body systems can provide relevant information
on PDFs at large $x_{Bj}$, while
it is known since a long time that its particular spin structure
suggests the use of $^3$He as an effective polarized
neutron target \cite{friar}.
Polarized $^3$He will be therefore the first candidate
for experiments aimed at the study of spin-dependent
GPDs in the free neutron, to unveil details of its angular momentum
content. 

In this talk, an Impulse Approximation (IA)
calculation of the quark unpolarized GPD $H_q^3$ of
$^3$He is reviewed. The calculation is 
fully described in \cite{io},
where the reader can find all the formalism, skipped
here.  
In \cite{prd}, for any spin $1/2$ hadron target
made of three spin $1/2$ constituents,
a convolution formula is obtained
for $H_q^3$, in terms of $H_q^N$,
the GPD of the internal particle, and 
a non diagonal spectral function $P_N^3(\vec p, \vec p + \vec \Delta)$:
\begin{eqnarray}
H_{q}^3(x,\xi,\Delta^2) =  
\sum_N \int_x^1 { dz \over z}
h_N^3(z, \xi ,\Delta^2 ) 
H_q^N \left( {x \over z},
{\xi \over z},\Delta^2 \right)~,
\label{main}
\end{eqnarray}
where 
\begin{equation}
h_N^3(z, \xi ,\Delta^2 ) =  
\int d E
\int d \vec p
\, P_N^3(\vec p, \vec p + \vec \Delta) 
\delta \left( z + \xi  - { 2 p^+ \over (P+ P')^+ } \right)~.
\label{hq0}
\end{equation}
Formally the above equation fulfills the theoretical constraints of
GPDs \cite{prd}
and it has been
numerically evaluated for $^3$He in \cite{io},
using a realistic  $P_N^3(\vec p, \vec p + \vec \Delta)$,
so that Fermi motion and binding effects are rigorously estimated.
In particular, for its evaluation
use has been made of wave functions overlaps 
evaluated in \cite{gema} by means of the AV18 NN interaction.
The proposed scheme is valid for $\Delta^2 \ll Q^2,M^2$
and despite of this it permits to
calculate GPDs in the kinematical range relevant to
the coherent, no break-up channel of deep exclusive processes off $^3$He.
In fact, the latter channel is the most interesting one for its 
theoretical implications, but it can be hardly observed at
large $\Delta^2$, due to the vanishing cross section.
The nuclear GPDs obtained here
are a prerequisite for any calculation of observables 
in coherent DVCS off $^3$He, 
although they cannot be compared with existing
data. Thus, the main result of this investigation
is not the size and shape of the obtained $H_q^3$ for $^3$He,
but the size and nature of nuclear effects on it.
This will permit to test directly, for the
$^3$He  target at least, the accuracy of prescriptions
which have been proposed to estimate nuclear GPDs \cite{poly},
providing a useful tool for the planning of future experiments
and for their correct interpretation.

Nuclear effects are found to be
larger than in the forward case 
($\Delta^2=0,\xi=0$)
and increasing with $\Delta^2$ and $\xi$.
They are also flavor dependent, being more important
for the flavor $d$ than for the flavor $u$.
They also depend
on the used realistic nuclear potential.
Besides, it is found that 
nuclear GPDs cannot be factorized into a $\Delta^2$-dependent
and a $\Delta^2$-independent term, as suggested
in prescriptions proposed for finite nuclei.

An illustration of the size and relevance of nuclear
effects is given in Fig. 1, where it is shown the ratio
$R^{(0)}$ 
(see \cite{io} for a detailed definition)
of the nuclear to nucleon GPDs $H_q$,
corresponding to the flavor $d$.
Such a ratio would be one if there were no nuclear effects.
It is clearly seen from the figure that nuclear effects
increase with $\xi$, $\Delta^2$ and that they
depend on the choice of the NN potential, at variance
with what happens in the forward case.

A detailed analysis of DVCS off $^3$He, with estimates
of observables, such as
cross-sections or spin asymmetries, is in progress. 

\begin{figure}[hb]
\includegraphics[width=0.49\textwidth]{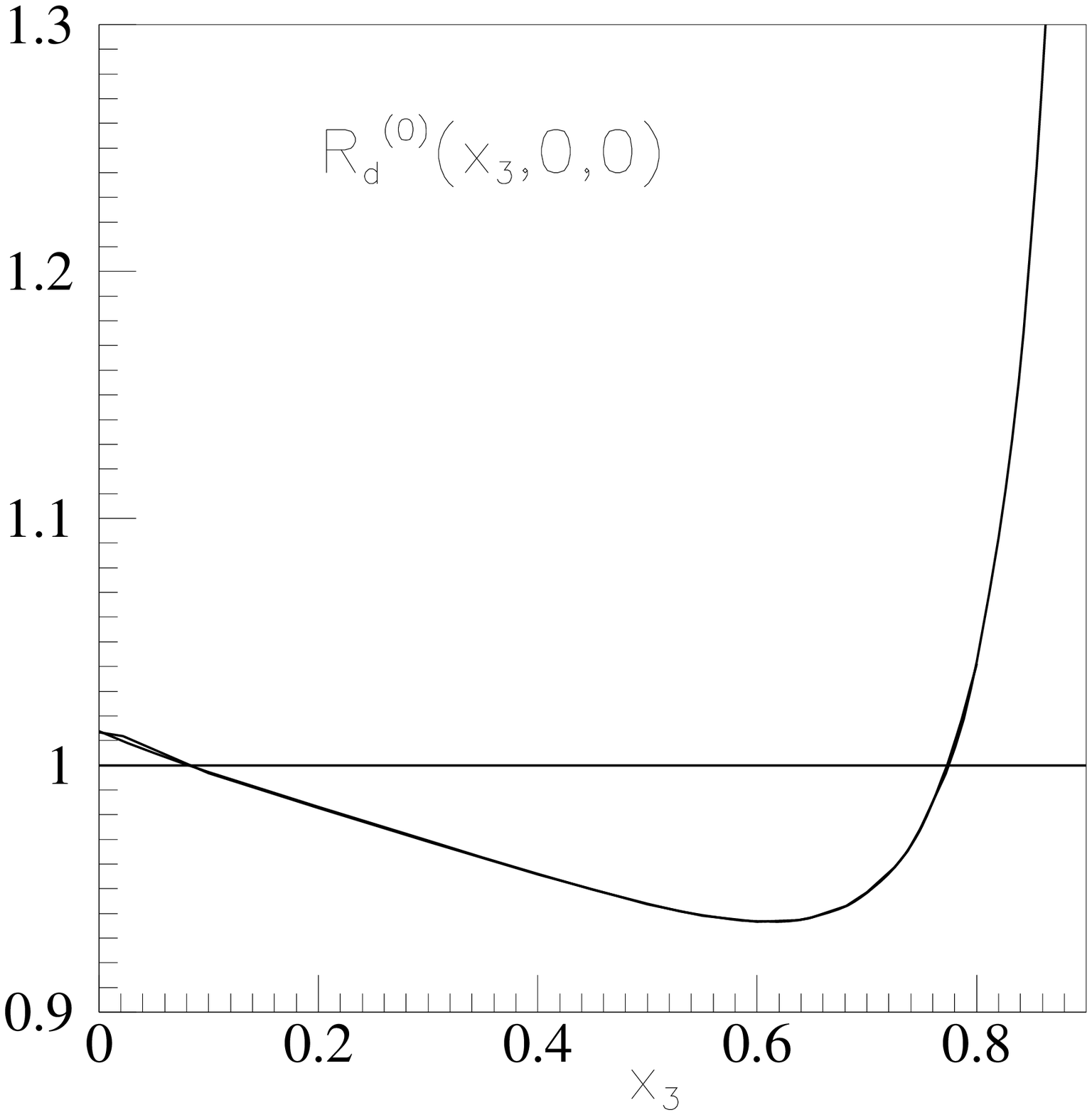}
\includegraphics[width=0.49\textwidth]{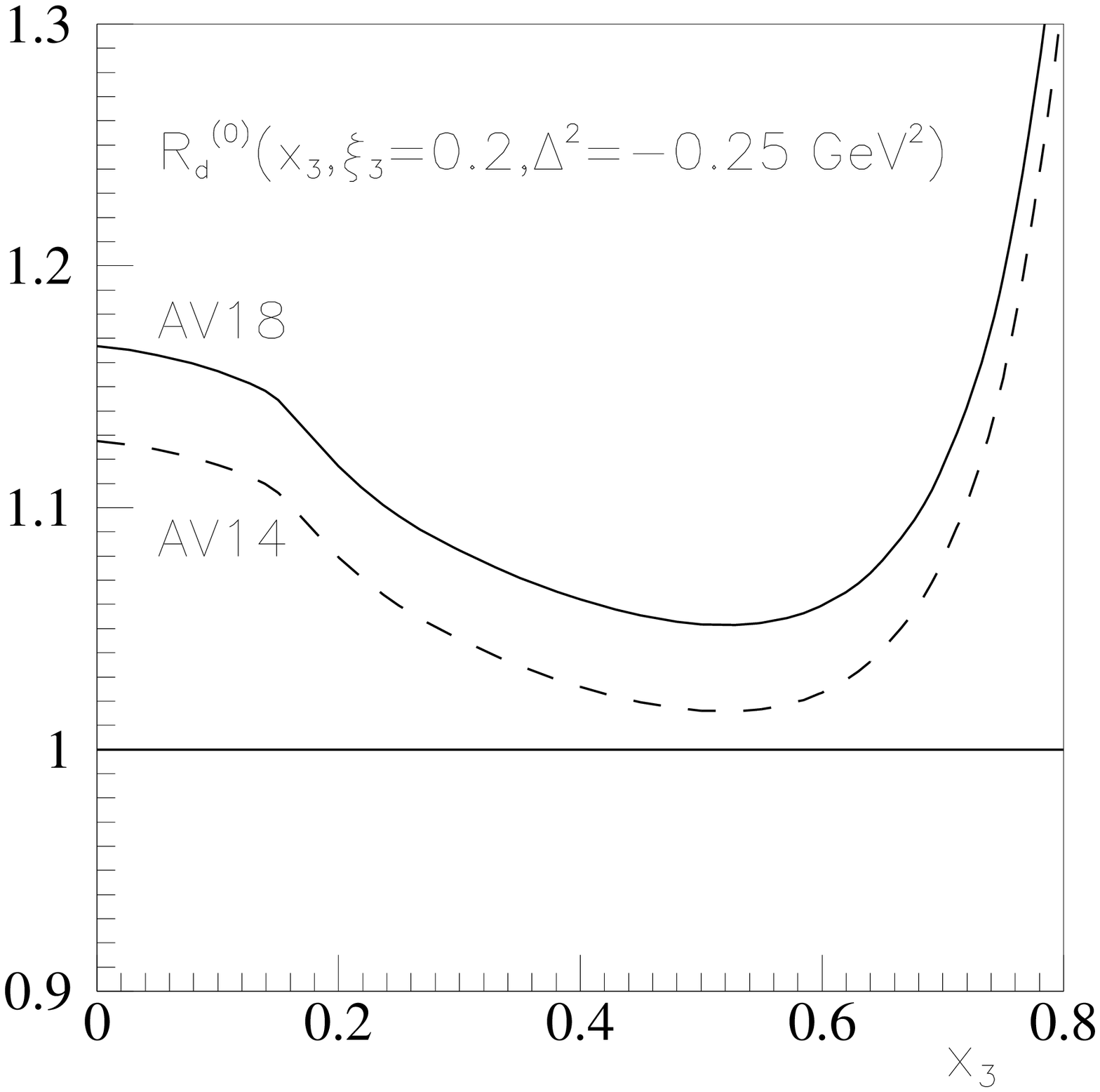}
\caption{Left panel: the ratio $R^{(0)}$, for the
$d$ flavor, in the forward limit $\Delta^2 = 0, \xi=0$, calculated
by means of the AV18 (full line) and AV14
(dashed line) interactions, as 
a function of $x_3 = 3 x$.
The results obtained with the different potentials are
not distinguishable.
Right panel: the same
as in the left panel, but at $\Delta^2=-0.25$ Ge$V^2$ 
and $\xi_3 = 3 \xi = 0.2$. The results are now
clearly distinguishable.}
\end{figure}

\vskip -20 cm

%%%%%%%%%%%%%%%%%%%%%%%%%%%%%%%%%%%%%%%%%%%%%%%%
%% BACKMATTER
%%%%%%%%%%%%%%%%%%%%%%%%%%%%%%%%%%%%%%%%%%%%%%%%

\begin{theacknowledgments}
I thank the Department of Physics of the Valencia University,
where part of this work was done, for a warm hospitality
and financial support. This work is supported in part by
the Italian MIUR through the PRIN Theoretical Studies of the
Nucleus and the Many Body Systems. 
\end{theacknowledgments}

%% If the bibliography is produced without BibTeX comment out th

\bibliographystyle{aipproc}   % if natbib is available

%\bibliography{sample}

\begin{thebibliography}{99}
\bibitem{first} D. M\"uller et al.
Fortsch. Phys. 42, 101 (1994); 
X. Ji, Phys. Rev. Lett. 78, 610 (1997);
A. Radyushkin, Phys. Lett. B 385, 333 (1996);
Phys. Rev. D 56, 5524 (1997).
\bibitem{dpr} M. Diehl, Phys. Rept. 388, 41 (2003).
\bibitem{cano1} E.R. Berger et al., Phys. Rev. Lett. 87, 142302 (2001);
F. Cano and B. Pire, 
Eur. Phys. J. A19, 423 (2004).
\bibitem{poly} M.V. Polyakov, Phys. Lett. B 555, 57 (2003);
V. Guzey and M.I. Strikman, Phys. Rev. C 68, 015204 (2003);
A. Kirchner and D. M\"uller, Eur. Phys. J. C 32, 347 (2003);
A. Freund and M.I. Strikman,
Phys.Rev.C 69, 015203 (2004).
\bibitem{friar} J.L. Friar et al., Phys. Rev.C 42, 2310 (1990);
C. Ciofi degli Atti et al., Phys. Rev. C 48, 968 (1993).
\bibitem{io} S. Scopetta, Phys. Rev. C 70, 015205 (2004).
\bibitem{prd} S. Scopetta and V. Vento, Phys. Rev. D 69, 094004 (2004).
\bibitem{gema} A. Kievsky, E. Pace, G. Salm\`e, and M. Viviani,
Phys. Rev. C 56, 64 (1997). 

\end{thebibliography}

\IfFileExists{\jobname.bbl}{}
 {\typeout{}
  \typeout{******************************************}
  \typeout{** Please run "bibtex \jobname" to optain}
  \typeout{** the bibliography and then re-run LaTeX}
  \typeout{** twice to fix the references!}
  \typeout{******************************************}
  \typeout{}
 }

\end{document}